\documentclass{article}
\usepackage{graphicx}
\date{ }
\begin{document}

\title{Comment on ``Inconsistencies in Verlinde's emergent gravity''}
\author{Youngsub Yoon \\\emph{Dunsan-ro 201, Seo-gu} \\\emph{Daejeon, 35245 Korea}}
\maketitle

\begin{abstract}
In 2016, Erik Verlinde proposed a new theory of gravity called ``emergent gravity'' by using mathematical formulas used in the theory of elasticity. In 2017, De-Chang Dai and Dejan Stojkovic claimed to point out inconsistencies in Verlinde's emergent gravity. We point out that their claim was based on misunderstanding of the dictionary between emergent gravity and theory of elasticity. In addition, we propose a slightly different formula for Verlinde's emergent gravity.
\end{abstract}

\section{Introduction}
In 2011, Verlinde proposed ``entropic gravity'' which claims that gravity is an entropic force \cite{entropic}. He derived Newton's universal law of gravitation and Einstein's equations by assuming the area law for entropy. In 2016, he developed his idea further, and proposed ``emergent gravity,'' in which the volume law of entorpy for gravity is considered in very weak gravity regime \cite{Verlinde}. He linked Milgrom's constant with the Hubble constant by linking the problem of missing mass in galaxies with the acceleration of our Universe by noting that our Universe is very close to accelerating de Sitter space. Thereby, he successfully derived Tully-Fisher relation, an empirical relation in galaxy rotation curve. In particular, he used the mathematical formulas used in the theory of elasticity. In 2017, De-Chang Dai, and Dejan Stojkovic claimed to find some inconsistencies in Verlinde's emergent gravity \cite{Inconsistencies}. We point out their claim was based on misunderstanding of the dictionary between emergent gravity and theory of elasticity. As a bonus, we also point out that the total gravity must be given by $g=\sqrt{g_B^2+g_D^2}$ instead of Verlinde's $g=g_B+g_D$, where $g_B$ is the gravity due to visible (baryonic) matter and $g_D$ is the gravity due to apparent dark matter. The organization of this paper is as follows. In Section 2, we review Verlinde gravity. In Section 3, we review ``Inconsitencies in Verlinde's emergent gravity.'' In Section 4, we clarify their misunderstanding and suggest why $g=\sqrt{g_B^2+g_D^2}$ must be true. In Section 5, we conclude our paper.

\section{Verlinde gravity and theory of elasticity}
In this section, we review the analogy between theory of elasticity and Verlinde gravity by closely following Verlinde's original paper \cite{Verlinde}. 
In theory of elasticity, we have the displacement field $u_i$. The linear strain tensor is given by 
\begin{equation}
\epsilon_{ij}=\frac 12 (\nabla_i u_j +\nabla_j u_i)
\end{equation}
and the stress tensor is given by
\begin{equation}
\sigma_{ij}=\lambda \epsilon_{kk} \delta_{ij}+2\mu \epsilon_{ij}
\end{equation}    
where $\lambda$ and $\mu$ are so-called Lam\'e parameters.

In Verlinde's emergent gravity, the displacement field $u_i$ is given by
\begin{equation}
u_i=\frac{\Phi_B}{a_0}n_i\label{ui}
\end{equation}
where $-n_i$ is the direction of gravity, $a_0=cH_0$ and $\Phi_B$ is given by
\begin{equation}
h_{ij}=\delta_{ij}-2\Phi_B n_i n_j=\delta_{ij}-\frac{a_0}{c^2}(u_i n_j+n_i u_j)\label{hij}
\end{equation} 
where $h_{ij}$ is the spatial metric. Verlinde calls it ``Newtonian potential,'' but it is slightly different because it concerns the space space component of metric instead of the time time component of metric. By the way, Verlinde only considers the limit $a_0$ is fixed but $c$ goes to infinity, so that $h_{ij}$ remains very close to the flat metric. In other words, he only considers the non-relativistic limit.

Then, he considers the ADM mass as follows,
\begin{equation}
M=\frac{1}{16\pi G}\int_{S_\infty} (\nabla_j h_{ij}-\nabla_i h_{jj})dA_i\label{ADM}
\end{equation}
which yields
\begin{equation}
M=\frac{a_0}{8\pi G}\int_{S_\infty} (n_j \epsilon_{ij}-n_i \epsilon_{jj})dA_i
\end{equation}
upon substituing (\ref{hij}). If we multiply the left-handside by $a_0$ we get a quantity with the dimension of a force. Thus, we can express
\begin{equation}
Ma_0=\oint_{S_\infty} \sigma_{ij} n_j dA_i
\end{equation}
where
\begin{equation}
\sigma_{ij}=\frac{a_0^2}{8\pi G} (\epsilon_{ij}-\epsilon_{kk} \delta_{ij})\label{sigmaij}
\end{equation}
This is consistent with the fact that the gravitational waves are not longitudinal, as Lam\'e parameters so obtained give
\begin{equation}
\mu=\frac{a_0^2}{16\pi G},\qquad \lambda+2\mu=0
\end{equation}
which says that the velocity of pressure (i.e., longitudinal) waves, which depends on $\lambda+2\mu$, is zero. Also, this stress tensor is consistent with the elastic self energy
\begin{equation}
\frac 12 M \Phi_B=\frac 12 \oint_{S_\infty} \sigma_{ij} u_j dA_i=\frac 12 \int \epsilon_{ij} \sigma_{ij} dV
\end{equation}

Verlinde defines some new expressions. (In his paper, he obtained expressions for arbitary space-time dimension, but we will just put $d=4$.) He defined ``surface mass density''
\begin{equation}
\oint_S \Sigma_i dA_i=M 
\end{equation}
which satisfies
\begin{equation}
\Sigma_i=-\frac{g_i}{4\pi G}
\end{equation}
where $g_i$ is the gravity field. This yields the gravitational energy
\begin{equation}
U_{grav}=\frac 12 \int dV g_i \Sigma_i
\end{equation}
On the other side of the correspondence, we have the elastic energy
\begin{equation}
U_{elas}=\frac 12 \int \epsilon_{ij} \sigma_{ij} dV
\end{equation}
which agrees with Verlinde's correspondence (there is a sign mistake in his paper)
\begin{equation}
\epsilon_{ij} n_j=-g_i/a_0,\qquad \sigma_{ij} n_j=\Sigma_i a_0\label{Sigmaia0}
\end{equation}
We also have the deviatoric strain tensor, i.e., the traceless part of strain tensor
\begin{equation}
\epsilon'_{ij}=\epsilon_{ij}-\frac 13 \epsilon_{kk} \delta_{ij}
\end{equation}
and $\epsilon$ and $\sigma$ defined by
\begin{equation}
\epsilon'_{ij}n_j=\epsilon n_i,\qquad \sigma_{ij} n_j=\sigma n_i\label{epsilonsigma}
\end{equation}
By combining (\ref{Sigmaia0}) and (\ref{epsilonsigma}), Verlinde obtains
\begin{equation}
\Sigma=\frac{\sigma}{a_0}
\end{equation}
and by combining (\ref{sigmaij}) and (\ref{epsilonsigma}), Verlinde obtains
\begin{equation}
\Sigma=\frac{a_0}{8\pi G}\epsilon\label{Sigma}
\end{equation}

Then, he shows that the deviatoric part of the elastic energy is given by
\begin{equation}
\frac 12 \int_B \epsilon'_{ij} \sigma'_{ij} dV=\frac{a_0^2}{16\pi G} \oint_{\partial B} u_i dA_i\label{devielastic}
\end{equation}
from which he obtains
\begin{equation}
\int_B \epsilon^2 dV=\frac 23 \oint_{\partial B} u_i dA_i\label{epsilon^2}
\end{equation}
Using Stoke's theorem, (\ref{ui}) and (\ref{Sigma}), Verlinde finally obtains
\begin{equation}
\left(\frac{8\pi G}{a_0}\Sigma\right)^2=\frac 23 \nabla_i \left(\frac{\Phi_B}{a_0}n_i\right)
\end{equation}
Actually, on the above equation, instead of $\Sigma$, Verlinde writes $\Sigma_D$, which is ``apparent dark matter surface density,'' which causes additional gravity that is traditionally due to dark matter. He is certianly correct in writing so; in his paper, he wrote $\Sigma_D=(a_0/8\pi G)\epsilon$ for (\ref{Sigma}). Nevertheless, in this section, we expressed $\Sigma$ without this $D$ subscript for a reason that will be clear in Section 4.

Anyhow, Verlinde says that the total gravity is given by the addition of the gravity due to visible matter ($g_B$) and the gravity due to the apparent dark matter ($g_D$) as follows.
\begin{equation}
\vec g=\vec g_B+\vec g_D
\end{equation}

\section{Dai and Stojkovic's criticism}
Dai and Stojokovic note that $\epsilon\sim \nabla u$ scales as $1/r^2$ because $u$ scales as $1/r$. Thus, they argue that the gravitational field which is proportional to $\epsilon$, indeed scales as $1/r^2$ just as the Newtonian gravity, and criticizes Verlinde for forcing $\epsilon$ to scale as $1/r$.  

Then, they consider their own version of the way Verlinde tried to derive $\epsilon\sim 1/r$. Going further from Verlinde's logic as its basis, they argue that (\ref{ui}) must be replaced by
\begin{equation}
u_i=\frac{\Phi_B}{a_0} n_i +\vec{\delta}(x,y,z)\label{uiDai}
\end{equation}
where $\vec\delta(x,y,z)$ is the fluctuation caused by the non-uniform distribution of removed entropy.$\footnote{We did not explain the concept of removed entropy in this paper, but interested readers can read Verlinde's paper and Dai and Stojkovic's paper.}$ From this, one can get
\begin{equation}
\epsilon(r)=\frac{H}{r^2}+f(x,y,z)
\end{equation}
where $H$ is a constant that is not quite important for their discussion, while $f(x,y,z)$ is due to $\vec \delta(x,y,z)$.

Then, the left-hand side of (\ref{epsilon^2}) can be represented as 
\begin{equation}
\int \epsilon^2 dV=\int \frac{H^2}{r^4} dV+ 2\int \frac{H}{r^2} f(x,y,z) dV+\int f(x,y,z)^2 dV
\end{equation}
Here, the second term is canceled out on average, as the direction $\vec{\delta}(x,y,z)$, which causes $f(x,y,z)$ is quite random. Thus, 
\begin{equation}
\int \epsilon^2 dV\approx\int \frac{H^2}{r^4} dV+\int f(x,y,z)^2 dV
\end{equation}
Then, one gets
\begin{equation}
\epsilon(r)\approx \sqrt{\frac{H^2}{r^4}+f(x,y,z)^2}
\end{equation}
As $f(x,y,z)$ scales as $1/r$, they explain, the gravity indeed seems to fall as $1/r$ for large $r$ as Verlinde argued. However, they point out, the apparent dark matter surface density seems to scale as $1/r^2$ as
\begin{equation}
\Sigma_D=\frac{1}{A} \int \frac{a_0}{8\pi G} \epsilon(r) dA=\frac{1}{A}\frac{a_0}{8\pi G}\left(\frac{H}{r^2}+f(x,y,z)\right)dA\approx \frac{1}{A}\frac{a_0}{8\pi G}\int \frac{H}{r^2} dA
\end{equation}
where in the last step they used again the fact that $\int f(x,y,z) dA$ is suppressed.

\section{The clarification}
Let's check that the gravitational energy is equal to the elastic energy for non-Verlindian theory, i.e., the Newtonian theory. Of course, we do not need to check it, because it is already proven, but confirming that $a_0$ plays no role in the result will give us understanding. As checking that that $a_0$ doesn't play any role is the important part, we will only calculate how the various values scale without too much worrying about the exact factors.

First,
\begin{equation}
u\sim \frac{1}{a_0}\frac{GM}{r},\qquad \epsilon\sim \frac{1}{a_0}\frac{GM}{r^2}
\end{equation}
\begin{equation}
\sigma\sim \frac{a_0^2}{8\pi G}\epsilon\sim a_0\frac{M}{r^2}
\end{equation}
Thus, we obtain
\begin{equation}
U_{elas}=\frac 12 \int \epsilon_{ij} \sigma_{ij} dV\sim \frac{GM^2}{r_{\mathrm{min}}}\label{NewtonianU}
\end{equation}
where $r_{\mathrm{min}}$ is the size of the mass $M$. It's just the gravitational self energy of object with mass $M$.

On the other hand, the elastic energy in (\ref{devielastic}) is given by
\begin{equation}
\frac 12 \int_B\epsilon'_{ij}\sigma'_{ij} dV=\frac{a_0}{4}Mr\label{a04Mr}
\end{equation}
Therefore, this energy is the one responsible for the apparent dark matter, as it has $a_0$ factor. On the other hand, it doesn't include the energy due to the visible matter, i.e., the Newtonian gravitational energy, (\ref{NewtonianU}) because it doesn't have $G$ factor. Therefore, it is justified to include the subscript $D$ to write
\begin{equation}
\Sigma_D=\frac{a_0}{8\pi G}\epsilon
\end{equation}
as Verlinde did, since we are considering only the energy of apparent dark matter in (\ref{a04Mr}).

Now, it is clear what was wrong with Dai and Stojkovic. If they intended to calculate $\Sigma_D$, as long as they are only considering the elastic energy responsible for the apparent dark matter (i.e., the right-hand side of (\ref{devielastic})), they cannot include the delta function term in (\ref{uiDai}), because they must single out the effect of apparent dark matter in this expression. Perhaps, they could have included the delta function term, if they considered the sum of Newtonian gravitational energy and the gravitational energy due to apparent dark matter instead of the gravitational energy due to apparent dark matter on the right-hand side of (\ref{devielastic}).  Nevertheless, that was not what they did, and in such a case, the relevant value is not $\Sigma_D$ but $\Sigma$ that combines $\Sigma_D$ and $\Sigma_B$.

Let's see what we get, if we consider the sum of the two different source of gravitational energies. We obtain
\begin{equation}
\frac 12 \int \epsilon^T_{ij}\sigma^T_{ij}dV=\frac 12 \int \epsilon^B_{ij}\sigma^B_{ij}dV+\frac 12 \int \epsilon'^D_{ij}\sigma'^D_{ij}dV
\end{equation}
where the superscript $T$ denotes ``total'' as in ``total gravity'', $B$ denotes the visible (baryonic), Newtonian one and $D$ denotes the apparent dark matter one.

In practice, it would look like
\begin{equation}
\frac{1}{2}\int dV g_T\Sigma_T=\frac 12 \int dV g_B \Sigma_B+\frac 12 \int dV g_D \Sigma_D
\end{equation}
where the last term is due to the elastic energy given by the right-hand side of (\ref{devielastic}). Thus, we obtain (upon not writing the subscript $T$)
\begin{equation}
g^2=g_B^2+g_D^2
\end{equation}
as $\Sigma$s are proportional to $g$s. This is different from $g=g_B+g_D$ by Verlinde \cite{Verlinde}.

\section{Discussions and Conclusions}
In his paper \cite{Verlinde}, Verlinde calls the following regime the ``sub-Newtonian regime'' or the ``dark gravity regime''
\begin{equation}
\frac{8\pi G}{a_0}\frac{M}{A(r)}<1
\end{equation}
This is when Verlinde's emergent gravity deviates from the Newtonian gravity. However, we believe that Verlinde's emergent gravity deviates (slightly) from the Newtonian gravity even when the above criteria is not satisfied. If Verlinde's emergent gravity deviates from the Newtonian gravity only in the sub-Newtonian regime, his formula $\vec g=\vec g_B+\vec g_D$ will become exactly $\vec g=\vec g_B$ in the Newtonian regime, i.e. when ``$>$'' is satisfied in the above inequality. However, if one calculates $g_D$ when ``$=$'' is satisfied in the above inequality, $g_D$ is not zero. Since $\vec g$ cannot suddenly jump from $\vec g_B+\vec g_D$ to $\vec g_B$ when it transits from the sub-Newtonian regime to the Newtonian regime, it is easy to see that Verlinde's emergent gravity does work not only in the sub-Newtonian regime, but also in the Newtonian regime as well.

Our formula $g=\sqrt{g_B^2+g_D^2}$ doesn't have this problem. The transition between the sub-Newtonian regime and the Newtonian-regime happens continuously and smoothly. Moreover, there is another problem with $g=g_B+g_D$. It would mean that the gravitational energy is given by
\begin{eqnarray}
\frac 12 \int dV g \Sigma=\frac 12 \int dV (g_B+g_D)(\Sigma_B+\Sigma_D)\qquad~~~~~~~~~~~~~~~~~~~~~~\nonumber
\\=\frac 12 \int dV g_B \Sigma_B +\frac 12\int dV (g_B \Sigma_D+g_D\Sigma_B) +\frac 12 \int dVg_D \Sigma_D
\end{eqnarray} 
We see that, in such a case, we would need to account for the energy due to the cross term, which is not yet certainly known and is unlikely to be present.
\pagebreak


\begin{thebibliography}{9}
\bibitem{entropic} 
E.~P.~Verlinde,
``On the Origin of Gravity and the Laws of Newton,''
JHEP {\bf 1104}, 029 (2011)
doi:10.1007/JHEP04(2011)029
[arXiv:1001.0785 [hep-th]].

\bibitem{Verlinde} 
E.~P.~Verlinde,
``Emergent Gravity and the Dark Universe,''
SciPost Phys.\  {\bf 2}, no. 3, 016 (2017)
doi:10.21468/SciPostPhys.2.3.016
[arXiv:1611.02269 [hep-th]].

\bibitem{Inconsistencies} 
D.~C.~Dai and D.~Stojkovic,
``Inconsistencies in Verlinde’s emergent gravity,''
JHEP {\bf 1711}, 007 (2017)
doi:10.1007/JHEP11(2017)007
[arXiv:1710.00946 [gr-qc]].


\end{thebibliography}
\end{document}